\documentclass[12pt]{iopart}
\usepackage{graphics}
\usepackage{epsfig}
\usepackage{citesort}

\usepackage{a4wide}

\usepackage{latexsym}
\usepackage{epsfig}
\usepackage{amssymb}
\newcommand{\bea}{\begin{eqnarray}}
\newcommand{\eea}{\end{eqnarray}}
\newcommand{\be}{\begin{equation}}
\newcommand{\ee}{\end{equation}}

\newcommand{\mc}[1]{\mathcal{#1}}

\newcommand{\ph}{\phantom}

\begin{document}

\title{Vector Field Models of Inflation and Dark Energy}


\date{\today}

\author{Tomi Koivisto}
\mailto{T.Koivisto@thphys.uni-heidelberg.de}
\address{Institute for Theoretical Physics, University of Heidelberg, 69120 Heidelberg,Germany}
\author{David F. Mota}
\mailto{D.Mota@thphys.uni-heidelberg.de}
\address{Institute for Theoretical Physics, University of Heidelberg, 69120 Heidelberg,Germany}

\begin{abstract}
We consider several new classes of viable vector field alternatives to the 
inflaton and quintessence scalar fields. Spatial vector fields are shown 
to be compatible with the cosmological anisotropy bounds if only slightly
displaced from the potential minimum while dominant, or if driving an 
anisotropic expansion with nearly vanishing quadropole today. The Bianchi I 
model with a spatial field and an isotropic fluid is studied as a dynamical 
system, and several types of scaling solutions are found. On the other hand, time-like 
fields are automatically compatible with large-scale isotropy. We show 
that they can be dynamically important if non-minimal gravity couplings
are taken into account. As an example, we reconstruct a vector-Gauss-Bonnet model 
which generates the concordance model acceleration at late times and 
supports an inflationary epoch at high curvatures. The evolution of 
the vortical perturbations in these models is computed.
\end{abstract}

\maketitle
\newpage
\tableofcontents

\section{Introduction}

Most inflationary and dark energy models \cite{Copeland:2006wr} are based on scalar fields \cite{Wetterich:1987fm}. 
They, as conventional cosmological sources, are fully characterized by their equation of state and speed of sound. For a more general phenomenology, one
should also take into account possible imperfect properties of a fluid \cite{Maartens:1997sh,Hu:1998tj}. Indeed, it has been proposed that dark energy
might have such a characteristic \cite{Koivisto:2005mm}, and the possibilities to distinguish it observationally have been examined 
recently \cite{Ichiki:2007vn,Mota:2007sz}. In Ref.\cite{Koivisto:2007bp} we contemplated on the possibility of such a significant anisotropy: 
that the expansion rate would be better described as a direction-dependent quantity within the Bianchi I model. The phenomenological set-up there  
could describe viscous fluids, Yang-Mills fields, strings or cosmology with noncommutative infrared properties. It was found that significant anisotropy may be 
allowed in some cases, and that these models have potential to explain the anomalous cosmological 
observations \cite{Eriksen:2003db,Land:2005ad}. In the future, one could 
optimistically expect to detect signatures of possible early anisotropies in the polarization, 
and of possible late anisotropies in the supernova luminosity 
distance-relationship data.     

In the present study we take a complementary approach and rather than working within a framework of phenomenological parametrization, we study vector field 
theories with explicitly written Lagrangians as our starting point. 
In particular, we look for 1) solutions which may be compatible with 
symmetries of cosmology, and then 2) which of these solutions might be used to model inflation or dark energy. The resulting models can then be divided into 
two classes, those with space-like and those with time-like field components. Both cases then seem to violate Lorentz invariance. However, this violation
is already done when choosing the cosmological background, not when adding the vector. The cosmological metric picks up a direction for time. 
The vector fields we consider are then particular solutions of a Lorentz-invariant theory compatible with the broken space-time 
symmetry of the metric. The conceptual difference with the Aether \cite{jac} and
such theories arises from the unit-norm requirement, which is achieved
by adding a norm-fixing term into the Lagrangian that then forces the
field to be always time-like, even in vacuum. In the present study we
do not impose the unit-norm requirement. See \cite{Kostelecky:1988zi,Bluhm:2004ep,Bluhm:2008yt} about spontaneous Lorentz violation in string theory and gravity.

Space-like vector fields in cosmology have been used to model inflation by Ford and others 
\cite{Ford:1989me,Burd:1991ew}. We will therefore focus on the present study more on the dark energy era, where the qualitative differences 
are that matter cannot be neglected and that the cosmologies do not have to isotropize. For completeness we also include a coupling to matter. In particular,
we work within the Bianchi I model, and look for scaling solutions. This approach differs from previous studies which have been confined to the case of the
exact Friedmann-Lema\^{i}tre-Robertson-Walker (FLRW) symmetry. Space-like fields are compatible with this symmetry only in the case of a so called triad, which 
has three identical vector fields for each spatial direction \cite{Armendariz-Picon:2004pm}. This can be used to model dark energy 
\cite{Armendariz-Picon:2004pm,Wei:2006tn,Wei:2006gv}. Inflation has been also considered in the case that a large amount of randomly oriented vector fields 
result in a average FLRW spacetime \cite{Golovnev:2008cf}. We note also that a vector field with quintessence can be considered \cite{Kiselev:2004py}, and that 
adding nonlinear terms to the electrodynamic Lagrangian could cause the universe to accelerate \cite{Novello:2003kh}.

Time-like fields have acquired a lot of interest in cosmology \cite{lim,skordis,bek,li,amar,franc,li2,aether,tom}.
If the field is canonical and minimally coupled, the equation of motion for its time-component becomes trivial. Therefore one has to either couple 
the field non-minimally or add non-canonical terms (these modifications are equivalent in some cases). Within modified gravity models attempting 
to eliminate dark matter, vector field components have been promoted to a central role due to their ability to mimic the observed gravitational lensing 
phenomena from dark matter \cite{bek,Contaldi:2008iw,Kiselev:2006gc}. 

Recently non-minimally coupled time-like vector fields have been taken under 
consideration as dark energy candidates \cite{Boehmer:2007qa,Libanov:2007mq,Jimenez:2008au}. Even if the 
vector field does not drive the inflationary expansion, it might have a role in slowing the expansion rate or on the generating perturbations 
\cite{lim,Dimopoulos:2006ms,Kanno:2006ty,Yokoyama:2008xw,Dimopoulos:2008rf}. Recently the observed anomalies in the CMB have inspired a number of studies 
about perturbations in an anisotropic background \cite{Pereira:2007yy,Pitrou:2008gk,Gumrukcuoglu:2006xj,Gumrukcuoglu:2007bx}, generation of structure 
during anisotropic inflation \cite{ackerman,pullen,Ando:2007hc,easson,barrow1,barrow2,Bohmer:2007ut} and on the formation of structure 
in the presence of anisotropic sources \cite{Koivisto:2008ig,Dulaney:2008bp}. In the present study we 
focus on role of the vector fields as driving the background expansion, especially the accelerating expansion associated with dark energy and inflation. 
The evolution of the spin-1 type perturbation in such a case is determined and we find these modes are decaying.  

We study cases where the field is coupled to gravity non-minimally. These vector-tensor couplings can cause the field to effectively break the usual energy 
conditions. Therefore they could be applied to model non-singular cosmologies and to construct worm-holes \cite{PintoNeto:2008id}. 
If the photon has such couplings during inflation, large-scale magnetic fields could be generated 
\cite{Turner:1987bw,Bamba:2008ja,bertolami}. In the present universe,
the vector-tensor theories can be severely constrained by Solar system experiments \cite{1972ApJ...177..757W,Bailey:2006fd}. We consider in particular two classes of 
these 
theories: with a coupling to the Ricci scalar and to the Gauss-Bonnet invariant. The latter has special interest, since the Gauss-Bonnet term has 
theoretical and phenomenologically desirable properties. In particular, its presence in the action does not immediately imply existence of ghosts and 
severe conflicts with observational data on gravitation. Our results include that this model could generate both the inflation and the present acceleration of 
the universe. The outline of the paper is simple: in section \ref{space} we derive our results concerning space-like models, in section \ref{time}
those about time-like models, and in section \ref{conclusions} we conclude.

\section{Space-like Fields}
\label{space}

\subsection{Minimally Coupled Vectors}
\label{v_mini}

Consider the vector field action
\be \label{action_min}
S = \int d^4 x \sqrt{-g} \left[\frac{1}{16\pi G} R -\frac{1}{4}F_{\mu\nu}F^{\mu\nu} - V(A^2)\right],
\ee
where $F_{\mu\nu} = \partial_\mu A_\nu - \partial_\nu A_\mu$ and $A^2 = A_\mu A^\mu$.
The field equations are
\be \label{fe_min}
G_{\mu\nu} = 8\pi G\left( T^m_{\mu\nu} + T^A_{\mu\nu}\right)
\ee
where the energy-momentum tensor of the vector field follows by varying with respect to the metric
\be \label{vector_emt}
T^A_{\mu\nu} = F_{\mu\alpha}F_{\nu}^\alpha + 2V'(A^2)A_\mu A_\nu - \left( \frac{1}{4} F_{\alpha\beta} F^{\alpha\beta} + V(A^2) 
\right) g_{\mu\nu}.
\ee
The equations of motion for the four vector components can be written as
\be \label{v_eom}
\nabla_\mu F^{\mu\nu} - 2A^\nu V'(A^2) = 0.
\ee
These follow either by setting the divergence of (\ref{vector_emt}) to zero or setting the variation of (\ref{action_min}) with respect to
 the field to zero.

We want to study the consequences of such matter sources in a Bianchi I type universe. Such is a generalization of the FLRW case with three different 
expansion rates. Thus we allow for a possible anisotropy. Then the metric may be written as 
\bea 
ds^2 & = & -dt^2 +a^2(t)dx^2+b^2(t)dy^2 + c^2(t)dz^2 \\ & = & -dt^2 + \sum_{i=1}^3 (a_i x_i)^2, \label{metric}
\eea
where we have introduced an obvious labeling of the spatial coordinates and the metric components with Latin letters.  
Following this convention, we denote $A_\mu=(\phi,A_i)$. Then the components of the energy-momentum tensor may then be written as
\bea \label{ta}
T^A_{00} & = & \frac{1}{2}\sum_{i=1}^3\frac{1}{a^2_i}\dot{A}^2_i + V(A^2) + 2V'(A^2)\phi^2, \label{ta00}
\\
T^A_{0i} & =  & 2V'(A^2)\phi A_i,
\\
T^A_{ij} & = & -\dot{A}_i\dot{A}_j + 2V'(A^2)A_iA_j +
a_i a_j\left(\frac{1}{2}\sum_{k=1}^3\frac{1}{a^2_k}\dot{A}^2_k - V(A^2) \right)\delta_{ij}. \label{taij}
\eea
Because the symmetries of the metric (\ref{metric}) do not allow a velocity field $T_{0i}$, one immediately sees
that we are restricted to consider either a purely time-like vector, $A_\mu = (\phi,0)$, or a space-like vector
with $\phi=0$. However, the former case reduces to triviality since
the equation of motion (\ref{v_eom}) for the nonzero vector component states that $V'(A^2)\phi = 0$ (since one has $\nabla_\mu F^{\mu 0}=0$).
Therefore we will consider the space-like vector fields.
For them, the off-diagonal components of $T_{ik}$ should vanish. In principle, one could then consider nontrivial (space-like) vector fields which
satisfy $$-\dot{A}_i\dot{A}_j + 2V'(A^2)A_iA_j = 0.$$ Then the evolution of the vector field, $A_i(t)$, would be determined a priori without reference
to the other matter (which then might or might not be consistently added into the system). Such possibilities seem contrived and we exclude
them from our present considerations. Thus, we have found that only space-like vector fields parallel to one of the principal axis could be allowed.
For the FLRW metric all the diagonal components of $T^i_{\ph{i}j}$ should be equal, which then requires three vector fields, one in each coordinate 
direction, of exactly equal magnitudes, as the triads of Armendariz-Picon \cite{Armendariz-Picon:2004pm}. With the metric Eq.(\ref{metric}), the only 
constraint is that every vector field should be along a coordinate axis. Any system of these vector fields is thus {\it required} to have
an anisotropic equation of state, unless it reduces to the very special triad case.  Let us consider the case $A_1=A_3=0$, $A_2=B$.
We then define the additional dimensionless parameters
\bea \label{xy}
X \equiv \frac{\kappa\dot{B}}{b H}, \quad Y \equiv \frac{\kappa^2V}{H^2}, \quad Y_1 = 2\kappa^2\frac{V' B^2}{b^2 H^2},
\eea
where $H$ is the average expansion rate
\be
H \equiv \frac{1}{3}\left(\frac{\dot{a}}{a}+\frac{\dot{b}}{b} + \frac{\dot{c}}{c}\right).
\ee 
It is also useful to define the fractional difference between expansion rates as
\be \label{er}
R  \equiv \frac{1}{H}\left(\frac{\dot{a}}{a}-\frac{\dot{b}}{b}\right).
\ee
This is a simple quantification of anisotropy with immediate generalization to the invariant shear.
We can then make contact with the parametrization of \cite{Barrow:1997sy,Koivisto:2007bp} by noting that
\be
w = \frac{X^2-2Y}{X^2+2Y}, \quad \delta = \frac{-X^2+Y_1}{\frac{3}{2}X^2 + 3Y}.
\ee
where $w$ is the equation of state along $x$-direction, and $\delta$ is the difference of the equations of state along $y$ and $x$ directions.
Generalization to multiple vector fields is straightforward.

\subsection{Scaling Solutions with and without Matter Couplings}
\label{scaling}

In this subsection we consider space-like cosmological vector fields. Similar anisotropic inflation has been considered in the early universe 
\cite{Burd:1991ew,Ford:1989me}. Therefore we focus on the present study more on the dark energy era, where the qualitative differences are that
matter cannot be neglected and that the cosmologies do not have to
isotropize. 
Our main aim is to find scaling solutions. There are several reasons to be interested scaling solutions.
1) Simply enough, if they are attractors, they may describe the realistic dynamics of the field. In any case they tell us about the 
phase structure of a given model. 
2) Scaling solutions have also the possibility to alleviate the coincidence problem. If
          dark energy and dark matter had similar (even within an order of magnitude
          or two) energy densities in the past, it seems less discomforting that
          the energy densities are about equal just today. However, it may still remain unexplained 
          why the transition to acceleration begun just
          at small redshifts - this is even if the dark matter and dark energy would
          continue scaling in the future, i.e. they would always have (roughly) equal
          magnitudes. Still, the coincidence problem a is good motivation to study 
          scaling solutions \cite{brook,luca,Amendola:2000uh,Koivisto:2005nr,val1}.
3) In addition, the possibility of early dark energy, which the scaling solutions
          automatically incorporate, is phenomenologically interesting \cite{Doran:2006kp}. A model
          predicting the presence of a dark energy component in the earlier universe
          can already be strongly constrained, and it may feature new signatures
          which might be useful in distinguishing between alternative models of
          acceleration.
Scaling solutions are known to exist in FLRW universes for
specific forms of scalar field Lagrangians and these have been applied in attempts to address the coincidence problem. 
Such a property could be shared by space-like vector fields in the Bianchi I background. 
For completeness we also include a coupling to matter. However, the presence of a 
coupling is not a necessary condition for this, as will be seen below. 

Consider the action  
\be \label{action_min2}
S = \int d^4 x \sqrt{-g} \left[\frac{1}{16\pi G} R -\frac{1}{4}F_{\mu\nu}F^{\mu\nu} - V(A^2) + q(A^2)L_m\right],
\ee
where we have included a general matter Lagrangian $L_m$ and the coupling function $q(A^2)$. 
To proceed, we write down the complete evolution equations using the variables (\ref{xy}). 
The amount of matter can be now described by a generalization of the usual fractional energy density,
\be
\Omega_m \equiv \frac{8 \pi G \rho_m}{3H^2}
\ee
The Friedmann equation may then be written in the form
\be \label{friedmann}
3 = 3\Omega_m + \frac{1}{2}X^2 + Y + \frac{1}{3}R^2.
\ee
Our time variable is chosen to be a generalization of the e-fold number. The prime will denote a derivative with respect to 
that,
\be
y' \equiv \frac{3dy}{d\log{abc}}.
\ee
It is convenient to derive an average effective equation of state, which then describes the overall expansion rate 
of the universe.
\be \label{weff}
w_{eff} \equiv -\frac{2}{3}\frac{H'}{H}-1,
\ee
Finally, we define two dimensionless variables, 
\be
\beta \equiv \frac{\kappa B}{b}, \quad \hat{q} \equiv \frac{3}{\kappa^2}q'(A^2)\beta,
\ee
which are the  comoving field $\beta$ and the coupling function $\hat{q}$. The $\kappa = 1/(8\pi G)$.

The system of equations then becomes
\bea
X' & = & -\left( \frac{H'}{H}+2+\frac{2}{3}R\right) X - \frac{Y_1}{\beta} - 2\hat{q}\left(1-3w_m\right)\Omega_m, \label{xprime}
\\
Y' & = & -2\frac{H'}{H}Y + \frac{Y_1}{\beta}\left[X-\beta\left(1-\frac{2}{3}R\right)\right], \label{yprime}
\\
R' & = & -\left(3+\frac{H'}{H}\right)R + X^2 - Y_1, \label{rprime}
\\
\Omega_m' & = & -\Omega_m\left[2\frac{H'}{H} + 3\left(1+w_m\right) - \frac{2}{3}X\hat{q}\left(1-3w_m\right)\right]. \label{oprime}
\eea
For a conventional scaling solution one should have $X'=Y'=\Omega_m'=0$. It is easy to see that this makes sense only if
also $R'=0$. Then the derivative of the average Hubble rate is a constant, $2H'/H = -3(1+w_{eff})$, which is
related to the effective (average) equation of state for the universe, $w_{eff}$.
Since also $w$ and $w+\delta$
now both correspond to the dark energy equations of state, $\delta$ and thus $Y_1$ must be a constant as well. One
deduces that this could be satisfied only if either $$ \textrm{1)} \beta'=0,
\qquad \textrm{2)} Y \sim (A^2)^n \qquad \textrm{or} \qquad \textrm{3)} Y=Y_1=0.$$ Below in table 
\ref{tab} we list the results for each case, but before we also derive and discuss these in more detail.

\begin{itemize}

\item In the case 1) that the comoving field stays constant, the square bracket term in
Eq.(\ref{yprime}) vanishes (since, as one can readily check, the derivative of the comoving field is proportional to the 
square bracket term). One uses this fact to relate the comoving field to $X$ and $R$. Moreover, now either $Y=0$ or 
$H'=0$ because Eq.(\ref{yprime}) must be satisfied. The 
former possibility of vanishing potential goes under the case 3) which we consider below. The latter possibility would 
correspond to the case $w_{eff} = -1$, which in this regard resembles a de Sitter-like space. However, since Eq.(\ref{rprime}) 
sets now $$R= \frac{X^2-Y_1}{3},$$ the solution is anisotropic unless $X^2=Y_1$. From Eq.(\ref{oprime}) one sees that for a scaling 
solution with 
$\Omega_m \neq 0$, one would have to force a coupling to keep the average Hubble rate constant. Since $\hat{q}$ is a 
function of the comoving field, the coupling term is now a constant. Plugging this into Eq.(\ref{xprime}) only gives us an 
equivalent of the Hubble equation as a consistency check. Therefore we have a whole set of solutions, which can be 
parametrized by the constants $Y_1$ and $\hat{q}$. We summarize these solutions as
\bea \label{scal_d}
w_{eff} & = & -1, \nonumber \\  X & = & \frac{9(1+w_m)}{2\hat{q}(1-3w_m)}, \nonumber \\ 
Y  & = &  
\Big\{ 648\hat{q}^3\left(1 - 3w_m \right)^3 Y_1 +
      \beta \Big[ -6561\left( -1 + w_m \right) \left( 1 + w_m \right)^3  +   
\nonumber \\ & + & 
         162\hat{q}^2\left( 1 - 3w_m \right)^2\left( 1 + w_m \right) 
          \left( 9 - 27w_m + 4w_m Y_1 \right)  \nonumber \\ & - &
         16\hat{q}^4\left( 1 - 3w_m \right)^4\left(Y_1^2 -81\right)
         \Big]\Big\}
\frac{1}{432\beta\left( \hat{q} - 3\hat{q}w_m \right)^4},
\\ \nonumber R &=& \frac{4\hat{q}^2(1-3w)^2}{3}, \\
\Omega_m &=& 
 \frac{\beta\left(1 + w_m\right)
           \left( 
             4\hat{q}^2\left( 1 - 3w_m \right)^2\left(Y_1-9\right)
             -81\left( 1 + w_m \right)^2
             \right)  + 4\hat{q}^3\left(3w_m-1\right)^3Y_1}{8\beta\left(\hat{q} - 3\hat{q}w_m \right)^4}
.\nonumber
\eea
The expression for the most general case is complicated. However, one notices that this scaling solution is unlikely to be able to describe a realistic 
era between the inflation and dark energy domination, since it implies that $H$ is equal to a constant, 
and requires a coupling to matter, which can easily be problematic \footnote{A possible solution to this issue would be to use 
chameleon vector fields \cite{ann,doug1,doug2}.}. 
    
\item In the power-law potential case 2) one notes that $Y'=0$ is consistent with Eq.(\ref{yprime}) only if $X=0$ 
(unless we would be back into either of the two other cases). Then $w=-1$ and $\delta=2n/3$. The coupling then does not 
affect the matter scaling since the field is constant, and Eq.({\ref{oprime}) tells that $w_{eff}=w_m$, since we would like to consider 
scaling solution with nonzero $\Omega_m$. The constancy of $R$ requires, from Eq.(\ref{rprime}) that $$R 
= \frac{4nY}{w_m-1}.$$ Plugging this back into Eq.(\ref{yprime}) we get that
$$Y=9(w_m-1) \frac{2n-3(1+w_m)}{16n^2},$$ This should of course be positive, imposing a constraint $n<3/2$ if $w_m=0$.
The amount of matter one then finds from the Friedmann Eq.(\ref{friedmann}), is given by
$$\Omega_m =  \frac{3}{8}\left(2- \frac{3(1+w_m)}{n^2}+ \frac{3+w_m}{n}\right).$$ This may be positive for $w_m=0$ when $n < -(3+\sqrt{33})/4
\approx -2.19$ or $n > (-3+\sqrt{33})/4 \approx 0.69$. Thus we have found that there exists a scaling solution for
a vector coupled to matter, which may be described as
\bea \label{scal_v}
w_{eff} &=& w_m \nonumber \\  X &=& 0, \nonumber\\ Y &=& \frac{9(w_m-1)[2n-3(1+w_m)]}{16n^2}, 
\\ R &=& \frac{-(9-6n+9w_m)}{4n},
\nonumber
\\
\Omega_m &=& \frac{3\left(2n^2-3(1+w_m)+n(3+w_m)\right)}{8n^2}.\nonumber
\eea
However, since we should have $X'=X=0$, one may be concerned with Eq.(\ref{xprime}). We mean that the kinetic contribution to the total energy should
stay negligible, and thus terms in the RHS Eq.(\ref{xprime}) should somehow cancel. One option would  to be to introduce a coupling which
cancels the effect of the potential that would otherwise set the kinetic contribution $X$ varying: $\beta\hat{q} =
-nY/((1-3w_m)\Omega_m)$. This could be consistent only for a coupling of the specific form $q(A^2) \sim log(A^2)$. However, from a practical point of view we rather notice that $Y_1/\beta$  can be approximately neglected if the
field has run to large enough values. We have found that this approximate scaling behaviour indeed arises naturally and is an attractor for the system for $n<-2.2$. These considerations can be confirmed by numerical computation.
In Fig.(\ref{scaling_pic}) we show some models featuring the scaling solution  Eq.(\ref{scal_v}).

\begin{figure}[ht]
\begin{center}
\includegraphics[width=0.49\textwidth]{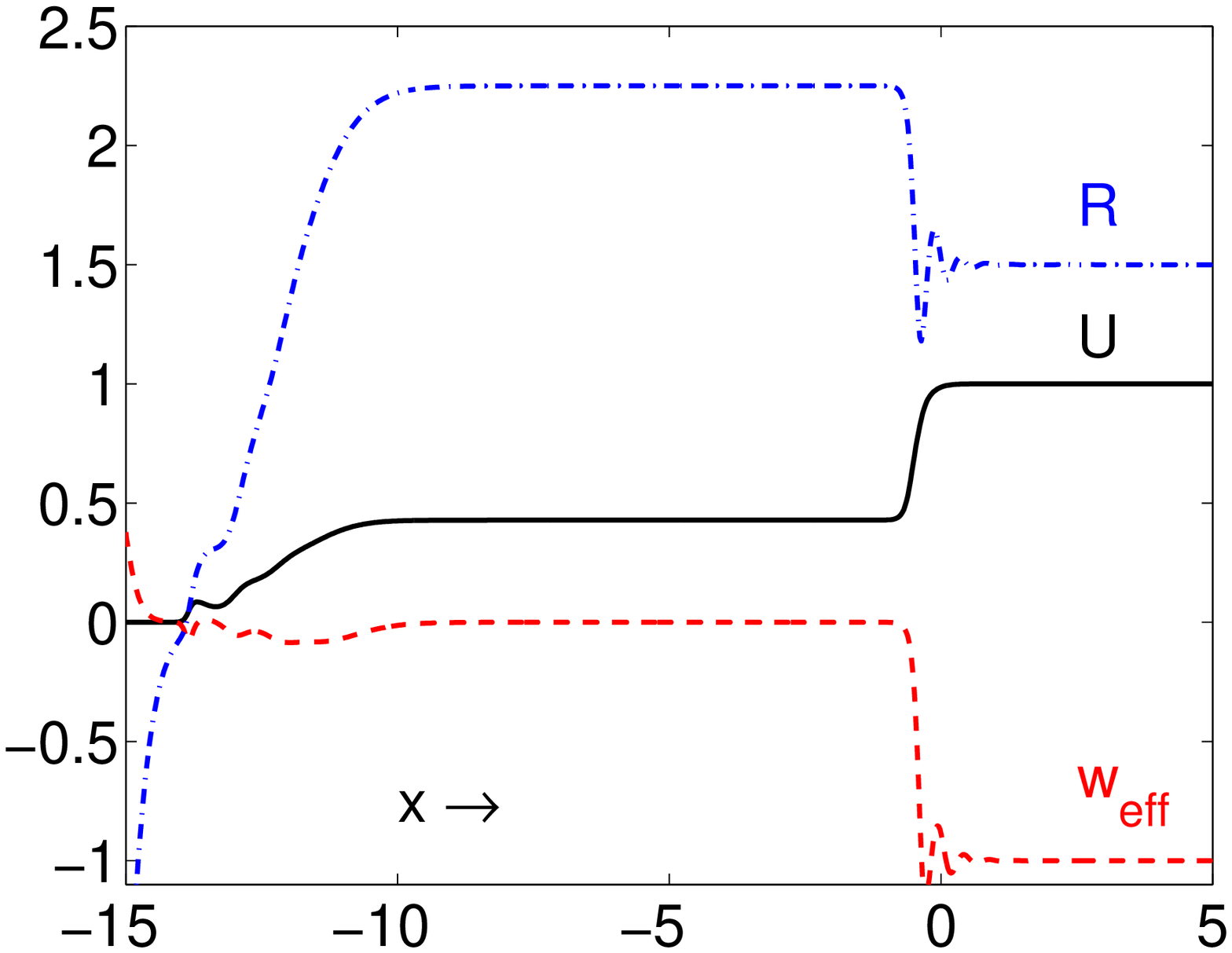}
\includegraphics[width=0.49\textwidth]{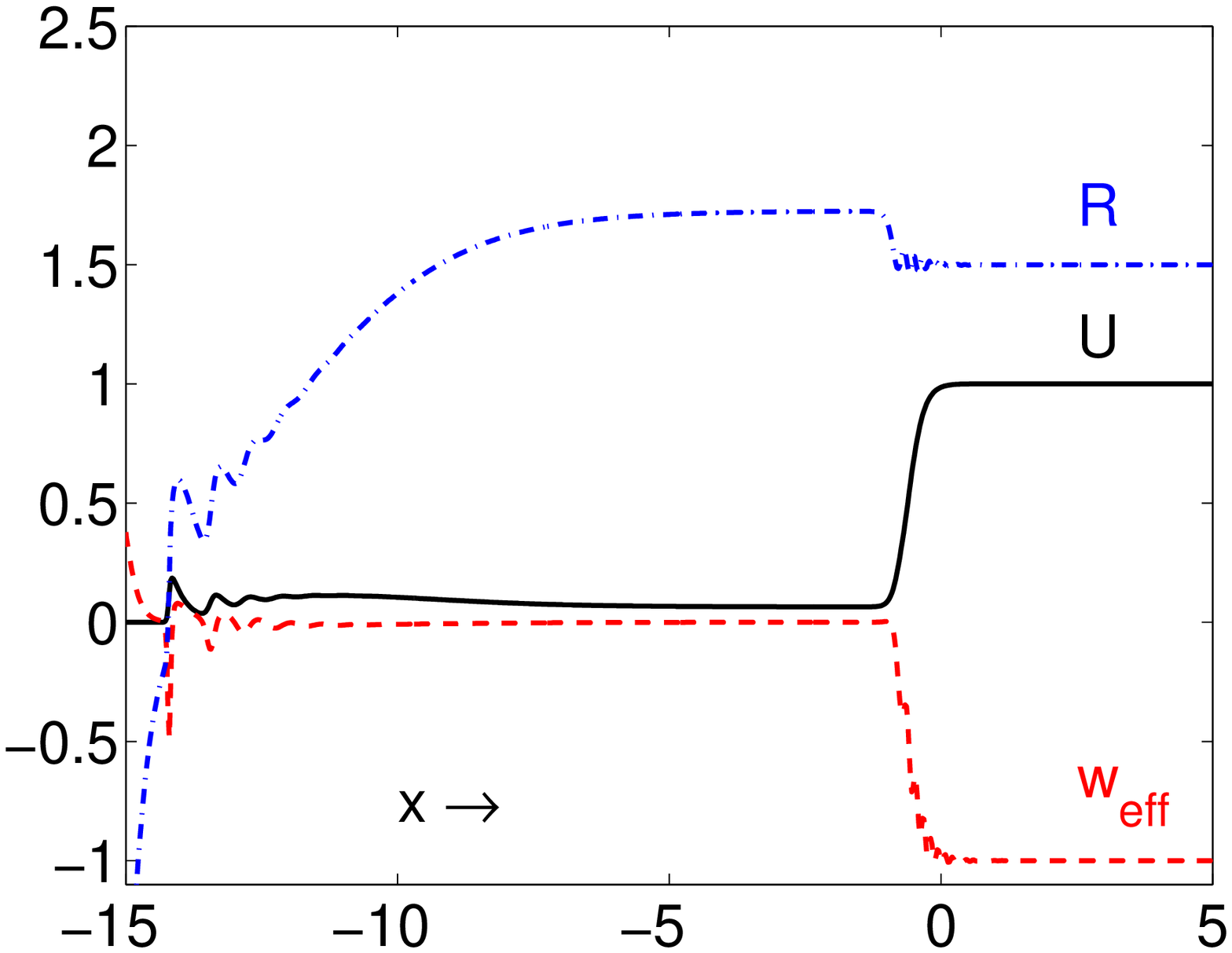}
\caption{\label{scaling_pic} Convergence to the scaling attractor Eq.(\ref{scal_v}) and the transition to the
accelerating era for minimally coupled space-like vector fields. The potential is of the form $V(x) =
V_+x^n + V_-x^{-n}$. In the LHS figure, $n=3$ and the RHS figure, $n=10$. The solid (black) lines are
the ratios $U$ of the vector to the total density, the dashed (red) lines are the average total equations of state
$w_{eff}$, and the dash-dotted (blue) lines are the expansion normalized shears $R$. The properties of the scaling
attractor depend only on $n$. The transition to acceleration depends on the scale $V_+$ which we have tuned such
that the vector dominance is taking place near the present time ($x=0$). }
\end{center}
\end{figure}

\item Let us finally consider the case 3) that the potential can be neglected. We see directly 
that for such a solution $w=1$ and $\delta=-2/3$. Again Eq.(\ref{oprime}) gives the effective equation of state,
$w_{eff} = w_m-2\hat{q}X(1-3w_m)/9$ which now may be influenced by the coupling. Then Eq.(\ref{rprime}) gives 
$$R= \frac{6X^2}{9+2\hat{q}X-3w_m(3+2\hat{q}X)}.$$ Inserting $\Omega_m=1-X^2/6-R^2/9$ in Eq.(\ref{xprime}) would now
give an expression for $X$, but this is generally quite complicated. Let us first look at the case that coupling vanishes. Then
\bea \label{scal_c}
w_{eff}&=&w_m,  \nonumber \\ \Omega_m &=& \frac{3(3-w_m)}{8}, \\ R&=&\frac{3(-1+3w_m)}{4}, \nonumber \\ X &=& \pm
\frac{3}{2\sqrt{2}}\sqrt{-1 + (4-3w_m)w_m}.\nonumber
\eea
The case of immediate concern to us is a background of dust, $w_m=0$, but there this solution does not exist since
$X$ should of course be real. Thus we are led to consider a case that the field is nonminimally coupled to dust,
$w_m=0$, $\hat{q} \neq 0$. This requires that the coupling function $\hat{q}$ is constant, which is possible, if
$\beta$ is not a constant, only for the special form of the coupling $q(A^2) \sim \sqrt{(A^2)}$. We again dismiss
this possibility.}

\end{itemize}

\begin{center}
\begin{table}
\begin{tabular}{|c|c|c|c|c|c|c|}
\hline
Solution & Equations      & $w_{eff}$ & $w$  & $\delta$ & Existence  & $R$  \\
\hline
Comoving field & (\ref{scal_d}) & $-1$      & $w(\beta,\dots)$  & $ w(\beta,\dots)  $ & Needs coupling &
$4\hat{q}^2/3$ \\
\hline
Potential driven & (\ref{scal_v}) & $w_m$     & $-1$ & $1/3$    & Large field   & $\frac{3}{2}-\frac{9}{4n}$\\
\hline
Kinetically driven & (\ref{scal_c}) & $w_m$     & $1$  & $-2/3$    &Not for dust  & -  \\
\hline
\end{tabular}
\caption{\label{tab} Scaling solutions for space-like vector fields and matter. The last column on the right
shows the shear $R$ in the dust filled universe. All of these solution exhibit anisotropy. }
\end{table}
\end{center}

We focus on the most relevant case, where dust dominates and there are no nonminimal couplings. We show that the scaling solution
(\ref{scal_v}) is an attractor. For that purpose, we eliminate the $H'/H$ and $\Omega_m$ terms from the system (\ref{xprime}-\ref{oprime}) and
get
\bea
X'  & = & \frac{X}{12}\left(-6-8R+2R^2+X^2-6Y+4nY\right) + \mathcal{O}(\frac{1}{\beta}), \\
\eea
\bea
Y'  & = & \frac{Y}{6}\left(18+2R^2+X^2-6Y+4n(-3+2R+Y)\right) + \mathcal{O}(\frac{1}{\beta}), \\ 
\eea
\bea
R'  & + & \frac{1}{12}\left(2R^3 + 12(X^2-2nY)+R(-18+X^2-6Y+4nY)\right) + \mathcal{O}(\frac{1}{\beta}).
\eea 
We consider then areas where $\beta$ is large and terms proportional to its inverse become negligible. 
Then consider the expansion 
\be
{\bf X} = \bar{{\bf X}} \epsilon {\bf x} + \mathcal{O}(\epsilon^2) = (\bar{R}, \bar{X},\bar{Y}) +\epsilon (x,y,z)
+ \mathcal{O}(\epsilon^2) , 
\ee
where background marked with bars is given by the solution (\ref{scal_v}). Then $\dot{{\bf X}} = 0$ at the background order, and to first 
order in $\epsilon$ we get 
\be
\dot{{\bf x}} = {\bf M}{\bf x},
\ee
where the three eigenvalues of the matrix ${\bf M}$ are given by 
\bea
[M_1,M_2,M_3] = \Big[\frac{3}{2}\left(\frac{1}{n}-1\right), 
& - & \frac{3}{4}\left(1+\frac{1}{|n|}\sqrt{(4n-3)(n^2+n-3)}\right), \\
& - & \frac{3}{4}\left(1-\frac{1}{|n|}\sqrt{(4n-3)(n^2+n-3)}\right)\Big].
\eea
The real parts of the eigenvalues are always positive for $n < -\frac{1}{4}(3+\sqrt{33})$, so the point is a local sink. This confirms that the 
scaling solution is an attractor given that the exponent $n$ is negative enough.

We have thus systematically considered every possible way towards a scaling solution of a cosmological (space-like) vector field which might
be coupled to matter. Several possibilities were noticed, and of particular relevance is the scaling solution
seems to be Eq.(\ref{scal_v}), which is valid in the limit of large enough field values. We should however note that several kinds
of different behaviours, anisotropic or not, with matter domination $\Omega_m=1$, or with vanishing matter
contribution $\Omega_m=0$, were left this time without explicit treatment, since we were only looking for a scaling
universe. In addition, more interesting solutions probably exist, in which the vector field tracks the background
matter (in analogy to the scalar field tracker quintessence) without the tracking being exact (which is precisely
what we mean by scaling solution). We summarize the scaling solutions in table \ref{tab}.

\subsection{Examples of models satisfying the anisotropy constraints}

In previous study \cite{Koivisto:2008ig} it was shown that the magnitude of the quadropole resulting from anisotropic
background expansion is given by 
\be
\label{quad}
Q_2 = \frac{2}{5\sqrt{3}}e_*^2,
\ee
where $e_*^2$ is the eccentricity, 
\be \label{ellips}
e_*^2 = \left(\frac{a_*}{b_*}\right)^2-1,
\ee
which tells how ellipsoidal the last scattering surface (denoted by star) appears to us (the scale factors normalized to unity today).
The constraint on the magnitude of quadropole is then $Q_2 \lesssim  2.72 \,\, 10^{-5}$
However, it is in principle possible for arbitrarily anisotropic expansion to escape detection from CMB as long as the expansion rates evolve 
in such a way that $e_*$ remains zero. In other words, the quadrupole vanishes, if each scale factor has expanded - no matter how anisotropically - 
the same amount since the last scattering.

\begin{figure}[ht]
\begin{center}
\includegraphics[width=0.49\textwidth]{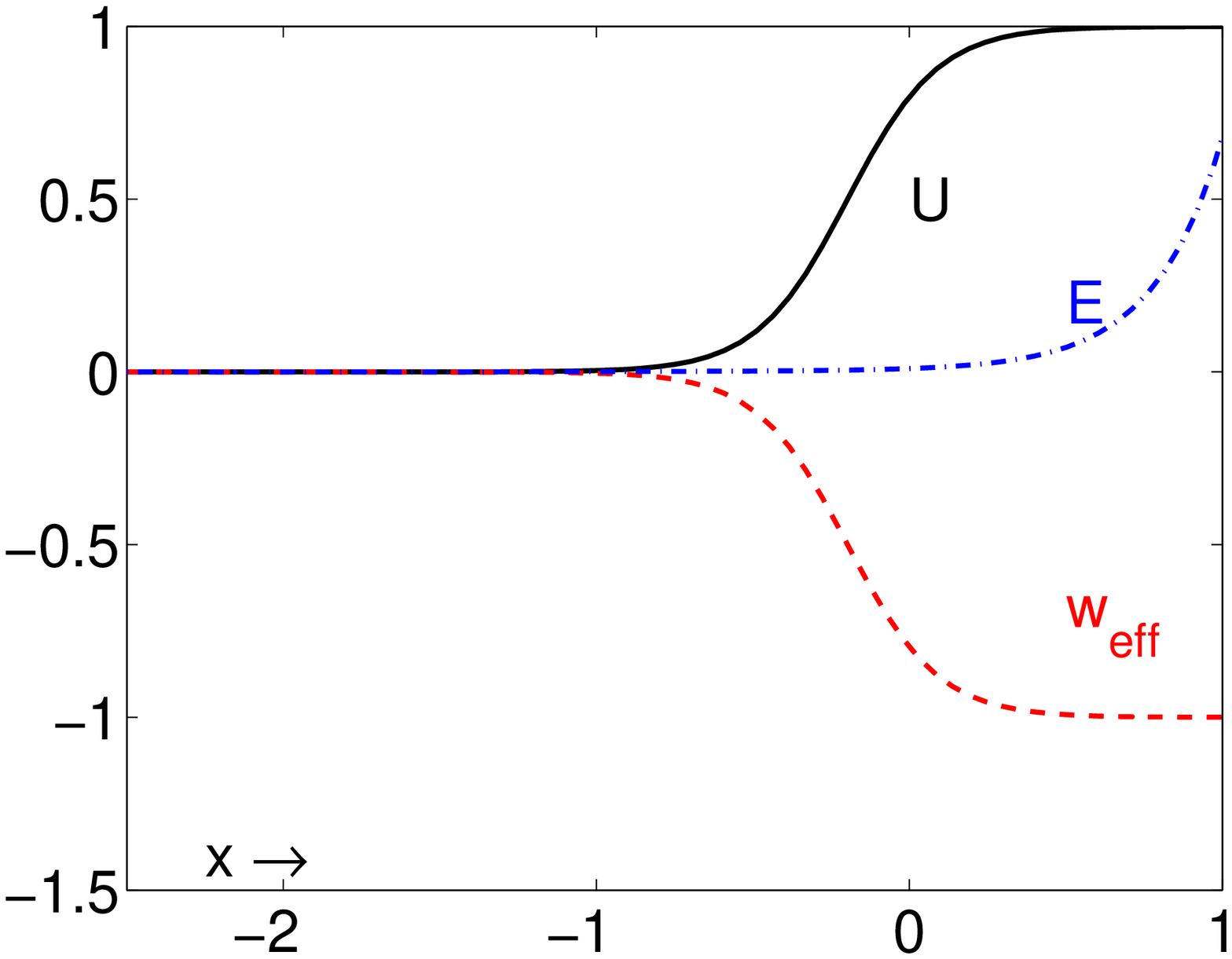}
\includegraphics[width=0.49\textwidth]{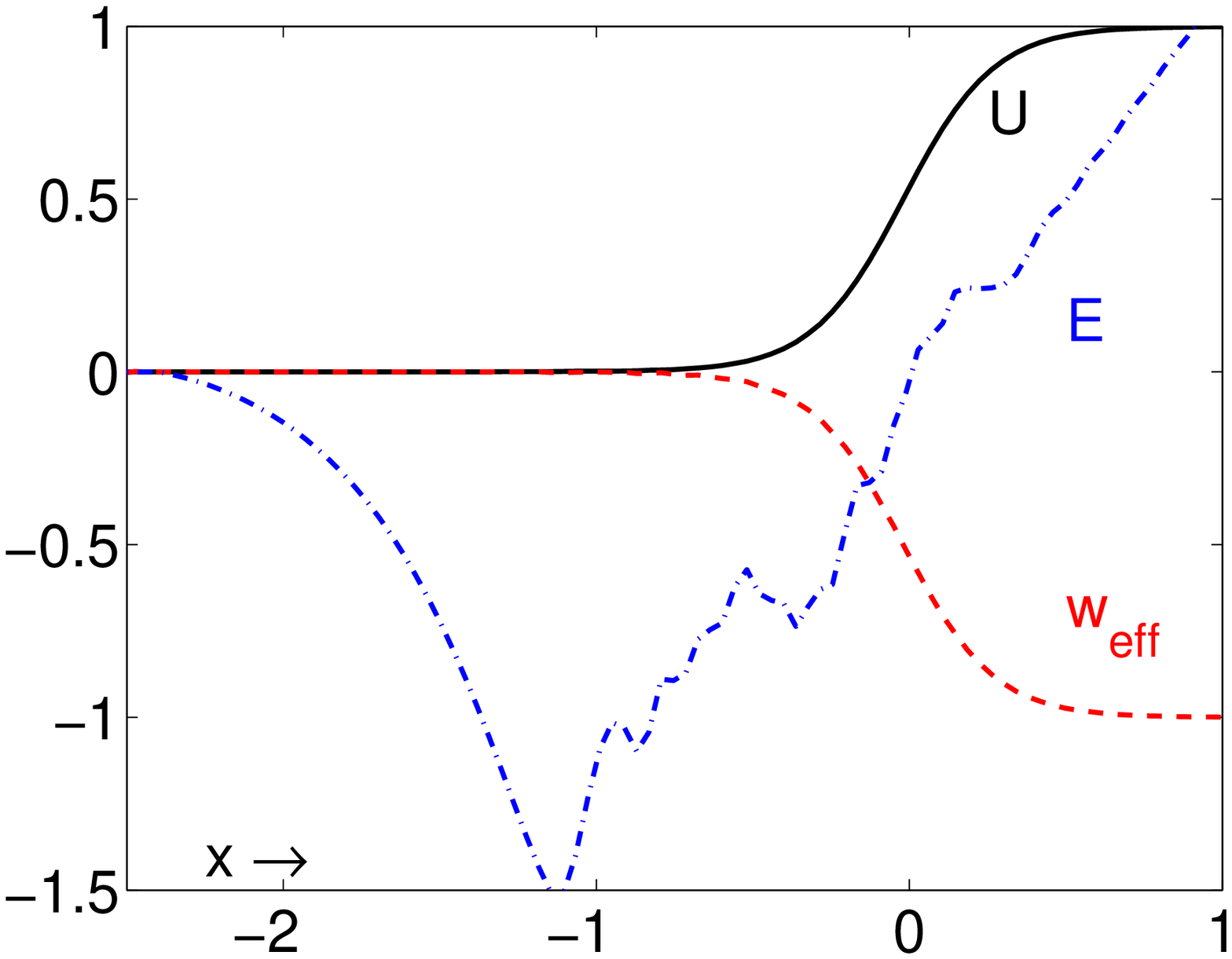}
\caption{\label{tuningpic} Two types of scenarios with a minimally coupled vector field that satisfy the CMBR quadrupole
constraint. The solid (black) lines are the vector field density fraction $U = \rho_A/(\rho_m+\rho_A)$, the dashed (red) lines are the effective
total equations of state of the universe $w_{eff}$, and the dash-dotted (blue) lines describe the evolution of
eccentricity, $E=500e_*^2$. In the LHS figure, the displacement of the field from the minimum of its potential is
important
only at late times and only then significant anisotropies begin to form. The potential here is an exponential,
$V(x) \sim e^{-x}$.
In the RHS figure, the potential is a double-power law $V(x) \sim x^{-4}+V_+ x$. The dynamics of the field is
such that though
there are significant anisotropies, the eccentricity at the present is nearly zero. Note that we have chosen both
the models in such
a way that they feature a recently begun acceleration. Thus these models are viable though finely tuned.}
\end{center}
\end{figure}

This is demonstrated explicitly with examples in Fig.(\ref{tuningpic}). In the left hand side (LHS) there is an example of an
exponential potential featuring large anisotropies at late times after the acceleration has begun. Although the anisotropy in such  case has been perturbatively 
small until today, such models could not be straightforwardly described within the standard approach of perturbing a FLRW metric.

Another example, in the right hand side (RHS) Fig.(\ref{tuningpic}), shows that the skewness of the universe, as defined by Eq.(\ref{er}), could be large, while
the eccentricity, as defined by Eq.(\ref{ellips}), might well vanish just at the present.  The latter is related to the
integral $\log(a/b) = \int R dx$, which might even be zero on average, if the shear $R$ happens to
oscillate suitably. This could naturally happen, for instance, if dark energy is a vector field slightly displaced from the
minimum of its potential. The model depicted in the RHS of Fig.(\ref{tuningpic}) exhibits this property and is constructed as a
simple, minimally coupled and one-component vector field introduced in the previous section \ref{v_mini}.
This is a fine-tuned model, but proves that there exist expansion histories exhibiting very large anisotropies with dynamics that leave the
quadrupole amplitude, Eq.(\ref{quad}), nearly or completely unchanged. 

The quadrupole constraint is applied here to constrain the overall amount of anisotropy, and we remark that the detailed anisotropy patterns which ensue 
within these constrains remain uncovered. 
The general evolution equations for the overdensities in the presence of anisotropic stresses have been presented \cite{Koivisto:2008ig}, and one may
observe that when the anisotropy is very small (as is forced by observations) the crucial modification 
comes from the anisotropic stress of the matter
source. It is this quantity which sets all the perturbation modes direction-dependent. This will have impact on the CMB spectra, in such a way that the 
correlations between adjacent $\ell$ modes become nonzero. Also unlike in the usual statistically isotropic case, different $m$-modes do not decouple. The 
detailed computation of anisotropies from the direction-dependent inhomogeneities \cite{Koivisto:2008ig} is left for forthcoming studies.

          To summarize, we constructed two classes of viable space-like vector dark energy models,
          described by the features: 
          1) The field evolves
          only very recently, and thus the impact of anisotropy on observations is
          strongly suppressed, or 2) There is earlier anisotropy, but in such a way that
          it's effect cancels today. Both cases require some fine tuning.
          Note however, that the quadrupole constraint suffices to estimate the
          amount of fine tuning. The reason is that only the quadropole is of the order
          of the background anisotropy in Bianchi I models. If the quadropole is not
          too large, the higher multipoles are of the observed order $10^{-5}$ \cite{Koivisto:2008ig}. 
          However, at the few lowest multipoles at least, there can appear some statistically
          anisotropic features. Thus we have found explicit viable mechanisms to 
          generate such anomalies in the late universe. 

For completeness, one may add that an approximate isotropy may also be achieved by 3) considering ''a triad'' of 
identical fields \cite{Armendariz-Picon:2004pm} or possibly 4) a huge number of randomly oriented fields \cite{Golovnev:2008cf}. In addition, let us note
the case 1) was mentioned already by Ford in the context of inflation: a very slowly rolling field resembles of course a cosmological constant.

\section{Time-like Fields}
\label{time}

\subsection{Vector Field Coupled to the Curvature}
\label{v_curv}

One might also allow nonminimal couplings of the vector and gravity. 
Couplings of a massless Maxwell vector field have been recently considered in an $f(R)$ model that could feature both the inflation and 
late-time acceleration of the universe, where the non-minimal coupling of the Maxwell field could then generate 
large-scale magnetic fields \cite{Bamba:2008ja,Bamba:2008xa}. Here we allow also a potential for the vector field, and do not couple the kinetic 
term of the field but add an interaction with the Ricci scalar $R$ and the field $A_\mu$,
\be \label{action_curv}
S = \int d^4 x \sqrt{-g} \left[\frac{1}{2}\left( \frac{1}{8\pi G} + \omega(A^2)\right) R - \frac{1}{4}F_{\mu\nu}F^{\mu\nu} - V(A^2) + L_m\right],
\ee
This is a special case of a vector-tensor theory \cite{jac,bek} satisfying the following three conditions\footnote{Whether the additional condition 
that the free-field energies are positive for both the metric and the vector can imposes further constraints on the form of $\omega$.}
1) the Lagrangian density is a four-scalar 2) the resulting theory is metric 3) there are no higher than second derivatives in the resulting field 
equations \cite{1972ApJ...177..757W}. 
The most 
general theory would allow also a coupling
of the form $A^\mu A^\nu R_{\mu\nu}$\footnote{Which in fact equals the kinetic term $A^\mu A^\nu R_{\mu\nu} \rightarrow (\nabla_\alpha A^\alpha)^2 
 - \nabla_\alpha A^\beta \nabla_\beta A^\alpha$, as is seen by using the geometric identity $[\nabla_\mu,\nabla_\nu]A^\alpha = 
R^\alpha_{\phantom{\alpha}\beta\mu\nu}A^\beta$ after a partial integration of the action.}. In the next subsection we however present a new model, 
which is of a quadratic form in the curvature invariants, but also seems to satisfy these requirements (depending on how to interpret 
condition 2)).

The contribution of the coupling term $\omega$ to the field equations can be presented as an effective energy-momentum tensor,
\be \label{fe_cur}
G_{\mu\nu} = 8\pi G\left( T^m_{\mu\nu} + T^A_{\mu\nu} + T^\omega_{\mu\nu}\right),
\ee
where $T^A_{\mu\nu}$ is given by Eq.(\ref{ta}) and $T^\omega_{\mu\nu}$ reads
\be
T_{\mu\nu}^\omega = -\omega G_{\mu\nu}+\left(\nabla_\mu \nabla_\nu -  g_{\mu\nu} \Box \right)\omega -
\omega' A_\mu A_\nu R.
\ee
To write the components explicitly for the line element (\ref{metric}), it is useful to introduce the
notation $A_{\mu} = (\phi,a_i\Lambda_i)$ for the field. We then find
\bea
T_{00}^\omega & = & -6H\left( {\bf \Lambda}\cdot {\bf \dot{\Lambda}}-\dot{\phi}\phi\right)\omega'
-\frac{1}{2}\left(9H^2-{\bf H}\cdot{\bf H}\right)\omega - \phi^2\omega' R, \label{tr00}
\\
T_{0i}^\omega & = & -\phi A_i \omega' R,
\\
T_{ij}^\omega & = & \label{trij}
2a_ia_j\left[ \left({\bf \Lambda}\cdot {\bf \dot{\Lambda}}-\dot{\phi}\phi\right)\left(3H-H_i\right)
+ {\bf \dot{\Lambda}}\cdot{\bf \dot{\Lambda}}+{\bf \ddot{\Lambda}}\cdot{\bf {\Lambda}}
-\dot\phi^2-\ddot{\phi}\phi\right]\omega' \delta_{ij}
\\ & - & 
a_ia_j\left(3\dot{H}+\frac{9}{2}H^2+\frac{1}{2}{\bf H}\cdot{\bf H} -\dot{H}_i-3H H_i\right)\delta_{ij}\omega
\\ & + & 4a^2\left({\bf \Lambda}\cdot {\bf \dot{\Lambda}}\right)\delta_{ij}\omega'' - A_iA_j\omega' R,\nonumber
\eea
where $R = 9H^2 + 6\dot{H} + {\bf H}\cdot{\bf H}$, and we have used the obvious notation for the 3-vectors ${\bf H}$,
${\bf H} = (H_1,H_2,H_3)$ and similarly for ${\bf \Lambda}$. 

The equation of motion for the time component of the field is 
\be 
\phi\left(2V'(A^2)-\omega(A^2)R\right) = 0.
\ee
The $G_{0i}$ component of the Einstein equations is identically satisfied if either the space-like or the time-like vector components vanish. 
The condition $G_{ij}=0$ then gives
\be \label{dc}
-\dot{A}_i\dot{A}_j + \left(2V'(A^2)-\omega(A^2)R\right)A_iA_j = 0.
\ee  
which translates into condition $\dot{A}_i\dot{A}_j=0$, which should be satisfied for all $i \neq j$. 
Given this condition and $\phi=0$, an arbitrary vector field is compatible with the Bianchi I metric. These requirements are identically satisfied
if we have only one evolving spatial vector component. This excludes the FLRW metric. 

In the remainder of this section we consider the special case of FLRW universe including a (solely) time-like 
field. Similar vector models have been recently 
considered by B\"ohmer and Harko \cite{Boehmer:2007qa} to successfully model the late acceleration of the universe while satisfying the the stringent 
Solar system constraints. Recently Jimenez and Maroto \cite{Jimenez:2008au} have also noted that a vector-like dark energy could avoid some of the 
fine-tuning problems present in scalar field quintessence type models. They showed a time-like vector field with a non-standard kinetic term could
 explain the SNIa data even without a potential. 

A time-like field does break the Lorentz invariance as well as the spatial fields considered in the previous subsection, since picking up the time coordinate 
introduces a preferred frame. 
However, in an FLRW universe one can treat the vector as an isotropic field whose  
time component resembles a nonstandard scalar field (there will be interesting changes when perturbations are taken into 
account). 

Summing the contributions from Eqs.(\ref{ta00}) and (\ref{tr00}) the energy density is
\be \label{r_rho}
\rho_\phi = 6H\phi\dot{\phi} -3H^2\omega + V,
\ee
where we have used the equation of motion $2V'=\omega' R$ to eliminate the last term in Eq.(\ref{tr00}). One notes that this solution does not
exist for the case of a massive vector field $V(x) = \frac{1}{2}m^2x$ linearly coupled to the curvature, $\omega(x) = \omega_0x$ except in the special
case of constant curvature. However, the more general case of nonlinear coupling and/or nonlinear potential these kind of dynamical solutions do exist.
The pressure of the field follows from Eqs.(\ref{taij}) and (\ref{trij}),
\be \label{r_p}
p_\phi = -V -2\left(2H\dot{\phi}\phi-\dot{\phi}^2-\ddot{\phi}\dot{\phi}\right)\omega' +
4\left(\dot{\phi}\phi\right)^2\omega'' + \left(2\frac{\ddot{a}}{a}+H^2\right)\omega.
\ee
One may check that the sum $\rho_\phi+p_\phi$ satisfies the consistency relation
$$\rho_\phi+p_\phi=-\dot{\rho}_\phi/(3H),$$ which follows from the covariant conservation of matter that applies now
since the $L_m$ is minimally coupled \cite{Koivisto:2005yk}. An interpretation is that the vector only changes the geometry. 
The conservation law may be 
written as
\be \label{r_sum}
\rho_\phi+p_\phi=\ddot{\omega}+H^3\left(\frac{\omega}{H^2}\right)^\bullet.
\ee
Let us first first look at de Sitter solutions assuming we may neglect $\rho_m$. We then set the Hubble parameter to a constant, $H_0$.
Firstly, one notes that an equation of motion is identically satisfied if the field is constant. Then the coupling follows from (\ref{r_sum}) inserted
into the Friedmann equation as $$\omega_0 = V/(3H_0^2) - (8\pi G)^{-1}.$$ Then, consistently, Eq.(\ref{r_p}) gives $p_\phi = -3H_0^2/(8\pi G)$. This holds
independently of the potential. One is thus able to mimic a cosmological constant with a time-like vector coupling, without having to have any potential present.
Secondly, one may also mimic a constant term even if the field is rolling. Then, however, the equation of motion gives the nontrivial constraint $V'=6H_0^2\omega'$
and one needs a potential $$V = 6H_0^2\omega + V_0.$$ If we for instance assume that the coupling has a power-law form, $$\omega(\phi) = \omega_1 \phi^\alpha + 
\omega_0,$$ then Eq.(\ref{r_sum}) tells us the field grows (or decreases, if $\alpha \gg -1$) exponentially with time, $\phi \sim e^{-H_0 t/\alpha}$. From 
Eq.(\ref{r_rho}) one may then deduce that if $\omega_1 = \alpha/2$ and that if $\omega_0=0$, then $V_0 \neq 0$. Similar results have been obtained 
in the unit-norm scalar-tensor-vector inflation model studied in Ref.\cite{Kanno:2006ty}, where it was noticed that a vector energy may stay constant even if the 
field is rolling down a potential, and on the other hand that such field can drive inflation even without a potential. 

Finally we will turn to late time universe where $\rho_m$ should be taken into account too. We look for a model which reproduces the background expansion of the 
$\Lambda$CDM model. The sum (\ref{r_sum}) should again vanish. The derivative of the Hubble rate would be given by $\dot{H}=\frac{1}{2}(\tilde{\Lambda}-3H^2)$, 
where $\tilde{\Lambda}=\Lambda/(8\pi G)$. Considering then the coupling as a function of the Hubble rate,
$\omega=\hat{\omega}(H)$, we find that its evolution equation:
\be \label{leg}
\frac{1}{2}(\tilde{\Lambda}-3H^2)\hat{\omega}''(H)-4H\hat{\omega}'(H)+2\hat{\omega}(H)=0
\ee
is of the Legendre form and thus has solutions in terms of the Legendre functions. Therefore, there
seems to be no simple form for the curvature-coupling function of time-like vector field mimicking the
$\Lambda$CDM expansion. From the (Legendre type) solution of Eq.(\ref{leg}) one could then deduce,
using the equation of motion, $2V'=\omega' R$, the form of the potential. But since these analytic forms would
be unappealing, we don't write them down here. The bottom line is that there exist solutions with nontrivial
vector field dynamics, with the energy density Eq.(\ref{r_rho}) residing in the field which however remains a constant.
The model can therefore share an identical background with the $\Lambda$CDM model at late times in addition to possibly driving an inflationary 
period at early times. In the following we will see, that with another kind of a curvature coupling, this can be achieved with polynomial 
forms for the coupling and potential functions. Such might appear more naturally in the low energy corrections to the gravitational action.
We leave the possible generation of magnetic fields in this model into future studies.

\subsection{A Vector - Gauss Bonnet Model}
\label{v_gb}

The Gauss-Bonnet topological invariant $R^2_{GB} = R^{\mu\nu\rho\sigma}R_{\mu\nu\rho\sigma} - 4R^{\mu\nu}R_{\mu\nu} + R^2$ can have
dynamical consequences in four dimensions if it interacts with the matter sector or has self-interactions. Dark energy 
cosmologies with self-interactions of $R^2_{GB}$ and with
scalar couplings have been considered previously \cite{Koivisto:2006xf,Koivisto:2006ai,Neupane:2006dp,li}.
Here we introduce a possible coupling to a vector and write the action as
\be \label{action_gb}
S = \int d^4 x \sqrt{-g} \left[\frac{1}{16\pi G}R - f(A^2)R^2_{GB} - \frac{1}{4}F_{\mu\nu}F^{\mu\nu} - V(A^2) + L_m\right].
\ee
In analogy with the discussions in the previous subsections, we now find an effective energy momentum tensor
\bea \label{gbtensor}
T^f_{\mu\nu} &=& -8\left[f_{;\alpha\beta}R_{\mu\phantom{\alpha\beta}\nu}^{\phantom{\mu}\alpha\beta}
+ \Box f R_{\mu\nu} -2 f_{\alpha ( \mu}R^\alpha_{\phantom{\alpha}\nu)}+\frac{1}{2}f_{;\mu\nu}R\right]
\\ &-&4\left[ 2f_{;\alpha\beta}R^{\alpha\beta}-\Box f R\right]g_{\mu\nu} + 2f'A_\mu A_\nu R^2_{GB}. \nonumber
\eea
To ease notation, we define $g^* \equiv -\dot{a}\dot{b}\dot{c}$. The determinant of the metric is
in our Bianchi I background $g = -abc$. We then have $R^2_{GB} = 8\dot{g}^*/g$. The
components of the tensor (\ref{gbtensor}) can be written as
\bea
T^f_{00} & = & 48f'\frac{g^*}{g}\left({\bf \Lambda}\cdot{\bf \dot{\Lambda}}-\phi\dot{\phi}\right)
+ 16 f' \frac{\dot{g}^*}{g}\phi^2, \label{tgb00}
\\
T^f_{oi} & = & 16f'\frac{\dot{g}^*}{g}\Lambda_ia_i\phi,
\\
T^f_{ij} & = & -16a_ia_j\left[\frac{1}{2}f'\frac{a_i}{g}\left(\frac{g^*}{a_i}\left({\bf \Lambda}\cdot{\bf\Lambda}
-\phi^2\right)^\bullet\right)^\bullet + 2f''\frac{a_i g^*}{\dot{a}_i g}
\left({\bf \Lambda}\cdot{\bf \dot{\Lambda}}-\phi\dot{\phi}\right)^2\right]\delta_{ij} \nonumber \\ &+&
16f'\frac{\dot{g}^*}{g}\Lambda_i\Lambda_j.  \label{tgbij}
\eea
Again the coexistence of space and time vector components is not generally possible, but in Bianchi I spacetimes one may
allow one space component to evolve simultaneously with the vector field.

Here we specialize to what may be the simplest possible model, a time-like vector field in FLRW spacetime.
Picking up the relevant terms from Eq.(\ref{ta00}) and Eq.(\ref{tgb00}), we get an energy density
\be \label{gb_rho}
\rho_\phi = -48H^3\dot{\phi}\phi f' + V,
\ee
where we have used the equation of motion $V'=-f'R^2_{GB}$ to drop a couple of terms. Again one might note that this solution does not
exist for the case of a massive vector field $V(x) = \frac{1}{2}m^2x$ linearly coupled to the Gauss-Bonnet invariant, $f(x) = f_0x$ except in the special 
case of constant $R^2_{GB}$. However, the more general case of nonlinear coupling and/or nonlinear potential these kind of dynamical solutions do exist.
The pressure of the field can be read from Eqs. (\ref{taij}) and (\ref{tgbij}),
\be
p_\phi = 16\left[2H\left(\dot{H}+H^2\right)\dot{\phi}\phi+H^2(\dot{\phi}^2+\ddot{\phi}\phi)\right]f' -
32H^2(\dot{\phi}\phi)^2f''
- V.
\ee
One can check that the sum
\be \label{gbsum}
\rho_\phi + p_\phi = 16H\left[\left(2\dot{H}-H^2\right)\dot{f}-\frac{1}{2}H\ddot{f}\right]
\ee
satisfies the consistency relation $$\rho_\phi+p_\phi=-\dot{\rho}_\phi/(3H).$$

In the remainder of this subsection we will attempt reconstructions of a coupling
and the potential in our action (\ref{action_gb}). For reconstruction schemes in other string-inspired cosmologies taking into account
(scalar) couplings of $R^2_{GB}$ term see Refs. \cite{Nojiri:2006je,Nojiri:2006be,Cognola:2006sp}.
de Sitter solutions in a vacuum would be relevant for the early inflation. Now it is clear that if the field is constant, one
would need a potential to achieve inflation. This is because as a topological invariant, the Gauss-Bonnet term cannot contribute to the dynamics if 
the coupling is a constant. However, if the field rolls down its potential, the coupling affects the dynamics, though possibly having a  
constant energy density with potential and thus yielding a de Sitter space as a solution, analogously to the Ricci-coupled case of the previous 
subsection. Let us pursue this first. 

The vanishing of the sum (\ref{gbsum}) requires $$f(t)=f_0 + f_1 e^{2H_0 t},$$ where $f_0$, $f_1$ and $H_0$ are 
constants. Since $f_0$ is classically 
irrelevant, we drop it. Then integrating the equation of motion gives us $$V(t) = V_0 - 24 H_0^4 f_1 e^{-2H_0 t}.$$ Here the constant $V_0$ corresponds 
effectively to a cosmological constant and we drop it also. If then we insert our result into the Friedmann equation, $H_0^2 = 8\pi G \rho_\phi/3$, where the 
energy density $\rho_\phi$ is given by Eq.(\ref{gb_rho}), we obtain an equation for the field as a function of time. The result is 
\be \label{gb_field}
\phi(t) = \frac{1}{4} + \frac{1}{256\pi G H_0^2 f_1}e^{2H_0 t}.
\ee
We then find that the potential and the coupling may have a simple inverse form,
\be \label{gb_inf}
V(A^2) = \frac{3H_0^2}{8\pi G}\frac{1}{\left(1-4|A|\right)}, \quad
f(A^2) = \frac{1}{64\pi G H_0^2}\frac{1}{\left(4|A|-1\right)}. 
\ee
There thus exists exact de Sitter solutions when the field is evolving. In fact the field is exponentially increasing as a function of the cosmic 
time $t$. Therefore additional terms, say positive powers of $|A|$, could
naturally terminate the de Sitter phase and end inflation. Hence it could originate a
possible graceful exit from inflation.    

Next we aim to find a vector-Gauss Bonnet model which produces, at late times, exactly the same background evolution as the
$\Lambda$CDM model. Since the vector then mimics a constant, the sum (\ref{gbsum}) should vanish. We want to have
$\dot{H} = (C-3H^2)$, where the constant $C=8 \pi \lambda$ is proportional to the corresponding cosmological constant $\lambda$. If we again think of the 
coupling as a function of the Hubble rate $\dot{f}=g(H)$, we can solve it as
\bea
&&(C-4H^2)g(H)-\frac{1}{4}(C-3H^2)g'(H) = 0 \nonumber \\ &\Rightarrow& \frac{dg}{g}=4\left(1-\frac{H^2}{C -
3H^2}\right)\frac{dH}{H}  \nonumber \\ &\
\Rightarrow& g(H) \sim H^4\left(C-3H^2\right)^{2/3}.
\eea
Furthermore, using the $\Lambda$CDM Hubble rate $$3H=(3H_0^2-C)e^{-3x}+C,$$
integrating once more we get the coupling as a function of $x$:
\bea \label{gpsolution}
f'(x) &\sim& H^3\left(C-3H^2\right)^{-2/3} \sim  e^{-2x}\left(1+\tilde{C}e^{-3x}\right) \nonumber
\\
& \Rightarrow &
f(x) = f_0 + f_1\left(220e^{-2x}+264\tilde{C}e^{-5x}+165\tilde{C}^2e^{-8x}+\tilde{C^3}e^{-11x}\right),
\eea
where $f_0$ and $f_1$ are integration constants and $\tilde{C}=3H_0^2/C-1$. The constant $f_0$ is dynamically irrelevant
and we drop it. Our model has two free functions: 
the potential $V$ and the coupling $f$. The solution Eq.(\ref{gpsolution}) tells how the coupling
has to evolve as the scale factor expands like in $\Lambda$CDM cosmology. How this corresponds to the evolution of the field
$\phi$, depends on the functional form of $f(A^2)$. We will choose its form as a polynomial in such
a way that $\phi \sim e^{-x/2}$:
\be \label{gb_coupl}
f(A^2) \sim \left[220 (-A^2) + 264\tilde{C}(-A^2)^{5/2} + 165\tilde{C}^2(-A^{2})^4 + \tilde{C^3}(-A^2)^{11/2}\right].
\ee
Then, from the equation of motion $V'=f'G$ we find that
\bea
V'(A^2) &\sim& \left[-2+\tilde{C}(-A^2)^{3/2}\right]\left[1+\tilde{C}(-A^2)^{3/2}\right]
\left[40+120\tilde{C}(-A^2)^{3/2} \right. \nonumber \\ &+& \left. 120\tilde{C}^2(-A^2)^{3}+\tilde{C}^3(-A^2)^{9/2}\right],
\eea
yielding the form for the potential as
\be \label{gb_pot}
V(A^2) \sim A^2\left[220+264\tilde{C}(-A^2)^{3/2} + 165\tilde{C}^2(-A^2)^3
+\tilde{C}^3(-A^2)^{9/2}\right],
\ee
where we have set an integration constant to zero since it would correspond to a cosmological constant.
We have thus found that a time component of a vector field coupled to the Gauss-Bonnet invariant
may generate 1) an early inflation epoch and 2) the $\Lambda$CDM background expansion without invocation 
of a cosmological constant or any scalar fields. This may happen when both the coupling and the 
potential are of a polynomial form, Eq.(\ref{gb_coupl},\ref{gb_pot}) in the case of dark energy and Eq.(\ref{gb_inf}) in the case of inflation.
These results seem quite interesting, keeping in mind that both vector fields and the Gauss-Bonnet 
term have crucial roles in fundamental theories. Suppose a theory gives corrections to the action in a form of power-law expansion
of the field strength $|A|$. Our results then imply that the negative powers could cause the early inflation
and the positive powers could be responsible for the present acceleration of the universe. 

\subsection{Vortical perturbations}

The conventional decomposition of the metric separates perturbations into scalar, vector and tensor quantities according to their
transformation properties under the spatial rotation group. Here we consider the vector-type (i.e. vortical) perturbations, and to avoid confusion we call
them spin-1 perturbations. In flat cosmology the metric, taking into account the spin-1 perturbations, can be specified without loss of 
generality by the line element
\be \label{metric-1}
ds^2 = a^2(\tau)\left[-d\tau^2 + {\bf B} \cdot d{\bf x} d\tau + d{\bf x}\cdot d{\bf x}\right].
\ee
We use here the conformal time $\tau = \int dt/a$. Generally the spatial sections would also have inhomogeneities, 
but we employ a gauge where they are flat. The shift vector ${\bf B}$ then fully characterizes the inhomogeneities. It is 
transverse, $\nabla \cdot {\cdot B} =0$. The vector field we write as
\be
A_\mu = a(\tau)[\varphi,{\bf L}]
\ee
where the transverse vector ${\bf L}$ is a perturbation. The conformal background field $\varphi = \phi(\tau)/a(\tau)$ is introduced for 
convenience. It follows that
\be
A^\mu = \frac{1}{a(\tau)}[-\varphi,{\bf L}+\frac{1}{2}{\bf B}]
\ee
and that $A^2 = -\varphi^2$, which is understandable since $A^2$ is of spin-0 type. We can also write up to linear order that
\be
a^2R=6(\mc{H}' + \mc{H}^2), \quad a^4 R^2_{GB} = 24\mc{H}^2\mc{H}',
\ee
where $\mc{H} \equiv a'/a$ is the conformal Hubble factor and prime denotes derivative with respect to conformal time. The equation of 
motion for ${\bf L}$ then reads
\be \label{spin1}
{\bf L}'' + 2\mc{H}{\bf L}' + \left[k^2 + (\mc{H}' + \mc{H}^2)(1+6\omega') + \frac{48}{a^2}f'\mc{H}^2\mc{H}' + 2a^2V' \right]{\bf L} = 0.
\ee
In this gauge the metric perturbations decouple from the evolution equation of the spin-1 type perturbation of the vector field. 
A wave-like propagation appears, unlike in the unit-norm models where the equation of motion instead becomes a first-order differential equation, 
whose solutions are diluting \cite{Lim:2004js}.
We are then 
able to deduce various properties of this perturbation. It is interesting to note that each component propagates like a scalar 
field fluctuation in the spatially flat gauge living in an effective potential given by the square bracket (excluding the gradient term). However, for the 
specific models considered in the last subsection the background equations motions reduce this to 
\be \label{spin2}
{\bf L}'' + 2\mc{H}{\bf L}' + \left[k^2 + (\mc{H}' + \mc{H}^2) \right]{\bf L} = 0.
\ee
Thus the evolution of the modes is independent of the detailed form of the coupling functions, and depends only on their effect to the background. 
As canonical scalar fields, the components of ${\bf L}$ propagate with light speed at small scales. This important fact establishes that
the spin-1 type perturbations are stable at small scales, quantum mechanically consistent (i.e. not ghost modes\footnote{To see this, the equation of 
motion itself is not enough. One has to also check that the overall sign of the vector Lagrangian in the action corresponds to the right sign of the
Hamiltonian, as now happens to be the case.}) and 
causal. For these models
we find that the modes decay during inflation also outside the horizon, though at slower rate than vector modes usually, ${\bf L} \sim a^{-1+\sqrt{10}/4}$. 
However, under an expansion with $w_{eff}>1/3$ the superhorizon vector modes become tachyonic. This kind of effect could amplify the vorticity seeds
during reheating. For radiation domination a constant mode exists. Given a regular primordial spectrum, vortical can result in large B-mode polarization 
signal at small scales $\ell \sim 2000$, which is distinguishable from the effects of tensor and gravitational lensing on polarization \cite{Lewis:2004kg}, and 
therefore possibly detectable with Planck. However, such effects might be pronounced in the present models. 

\section{Conclusions and Discussion}
\label{conclusions}

We investigated field theories where vector fields mediate non-gravitational interactions and drive 
the accelerated expansion of our universe, both during the early 
inflation and the 
present dark energy domination. The motivation for this investigation
comes from the frequent appearance vector fields and possibile Lorentz violations in fundamental theories, from the need to the test the robustness of 
the basic assumptions of cosmology, and from the hints of statistical anisotropy of our universe that several observations seem to suggest. 

In this study we considered the question whether such vector fields could have
an important  cosmological role instead of the more popular hypothetical scalar fields and their exotic variants. A usual reason to exclude
this otherwise reasonable possibility is that vector fields are incompatible with the established isotropy of the universe. However, space-like
fields could feature only small anisotropy, which some anomalous observations indeed hint could exist in the universe. Secondly, time-like
fields do not necessarily break isotropy, but they have non-trivial dynamics only if non-minimally interacting with gravity.
To summarize our results, we found for space-like fields that
\begin{itemize}
\item There exists for space-like fields (anisotropic) scaling solutions even without couplings to matter.
\item The cosmological bounds are satisfied in models of possibly large anisotropy if 1) the anisotropic era begins just recently or 2) the eccentricity 
cancels today. 
\end{itemize}  
We learned about (time-like) vector-tensor models that
\begin{itemize}
\item With nonminimal couplings to gravity, dynamical solutions for cosmological vector fields become naturally available. 
\item In particular, there are models with coupling to the Ricci or to the Gauss-Bonnet scalar that incorporate inflation and dark energy.
\end{itemize}  

In general, the amount of fine-tuning in vector dark energy models is similar to the amount of fine-tuning in scalar field models. As we have shown 
that there exists scaling attractors, the initial conditions for the fields and their velocities do not matter. One is then left to explain the energy
scales involved. Just like with scalar fields, by redefinitions and zero-point shiftings one may choose the numbers that go
into the Lagrangian, but the field turns out to be extremely light. As an example, consider the exponential potential, 
$V(A^2)=V_0e^{\kappa\lambda(A^2-A^2_0)}$.
If the potential energy is a significant contribution to the expansion rate today, the Friedmann equation tells that the potential is of the order 
$V \sim H^2_0/\kappa $. However, the potential scale $V_0$ can be set to whatever seems natural by choosing suitably the parameter 
$A^2_0$. The mass of the associated particle, $m_A \sim \sqrt{V'} \sim 10^{-33}$ eV, if the slope $\lambda$ is roughly of order one. For 
the power-law 
potential, $V(A^2) = V_0(\kappa A^2)^n$, one finds the exact relation
\be
m_A = \sqrt{-n}\left(\frac{V}{V_0}\right)^{\frac{1}{2}-\frac{1}{2n}}\sqrt{\kappa V_0},
\ee 
where $V$ is the value of the potential at a given time. If the time is nowadays and the scale of the potential is of order of the Planck 
mass, then the mass of the field is the Planck mass suppressed by a factor $10^{-50(1-1/n)}$, which for large inverse powers can be quite 
tiny. These considerations hold 
also even if the potential vanishes \cite{Jimenez:2008au}, since one may interpret that the mass effectively arises from the nonminimal coupling.
For the vector inflatons, one may find $TeV$ scale masses and potentials.

Thus, several cases of observationally viable and theoretically motivated vector fields in either FLRW or in a Bianchi I background were found. Those 
include models where again scaling solutions do exist for (space-like) vectors, with and without couplings to matter, and with given forms of the vector 
field potentials.
With matter and the vector field present, there are three kinds of scaling solutions, which were summarized in Table (\ref{tab}).
Allowing anisotropy of the dark energy equation of state could be useful in the quest for a realistic description of the present 
cosmological acceleration, since the exact symmetries of the FLRW metric exclude from the beginning for example these space-like vector 
fields. 

We also considered vector fields non-minimally coupled to gravity. We showed that a time-like vector coupled to
the Ricci scalar or to the Gauss-Bonnet invariant may generate both the inflation and the $\Lambda$CDM expansion. In the vector
Gauss-Bonnet model, the form of the potential and the coupling which mimic the concordance model background
expansion can be simple polynomial functions. Negative powers may drive
inflation and positive powers the late time acceleration. Amusingly, both the time-like field and the Gauss-Bonnet term become trivial
if the coupling is set to zero. Indeed, we have presented the first inflation models driven by time-like vector field. This may have
relevance to the problem of definition of the initial vacuum state of the universe. We considered the evolution of spin-1 type perturbations
showing that it is both causal and stable.

\section*{Acknowledgments} 

DFM acknowledges support from the A. Humboldt Foundation. 

\section*{References} 

\bibliography{birefs2}

\end{document}